\documentclass[sigconf,screen]{acmart}
\usepackage[nameinlink]{cleveref}
\usepackage{graphicx}
\usepackage{multirow}
\usepackage[table]{xcolor}
\usepackage{hhline}
\usepackage{tcolorbox}
\usepackage{makecell}
\usepackage{booktabs}
\usepackage{todonotes}

\title[LLM-Assisted Repository-Level Generation with Structured Spec-Driven Engineering]{LLM-Assisted Repository-Level Generation \\ with Structured Spec-Driven Engineering}

\author{Shuzhao Feng}
\orcid{0009-0001-0743-7219}
\affiliation{%
  \institution{McGill University}
  \city{Montreal}
  \state{Quebec}
  \country{Canada}
}
\email{shuzhao.feng@mail.mcgill.ca}

\author{Boqi Chen}
\orcid{0000-0002-1451-3603}
\affiliation{%
  \institution{University of Ottawa}
  \city{Ottawa}
  \state{Ontario}
  \country{Canada}
}
\email{boqi.chen@uottawa.ca}

\author{Brett H Meyer}
\orcid{0000-0002-6650-3298}
\affiliation{%
  \institution{McGill University}
  \city{Montreal}
  \state{Quebec}
  \country{Canada}
}
\email{brett.meyer@mcgill.ca}

\author{Gunter Mussbacher}
\orcid{0009-0006-8070-9184}
\affiliation{%
  \institution{McGill University}
  \city{Montreal}
  \state{Quebec}
  \country{Canada}
}
\email{gunter.mussbacher@mcgill.ca}

\copyrightyear{2026}
\acmYear{2026}
\setcopyright{cc}
\setcctype{by}
\acmConference[FSE Companion '26]{34th ACM Joint European Software Engineering Conference and Symposium on the Foundations of Software Engineering}{July 05--09, 2026}{Montreal, QC, Canada}
\acmBooktitle{34th ACM Joint European Software Engineering Conference and Symposium on the Foundations of Software Engineering (FSE Companion '26), July 05--09, 2026, Montreal, QC, Canada}
\acmDOI{10.1145/3803437.3805567}
\acmISBN{979-8-4007-2636-1/2026/07}

\date{January 2026}

\ccsdesc[500]{Software and its engineering~Automatic programming}
\ccsdesc[300]{Software and its engineering~Unified Modeling Language (UML)}
\ccsdesc[300]{Software and its engineering~Abstraction, modeling and modularity}
\ccsdesc[500]{Computing methodologies~Machine learning}
\ccsdesc[300]{Computing methodologies~Planning with abstraction and generalization}

\keywords{Large Language Models (LLMs), Code Generation, Spec-Driven Development, Artificial Intelligence, Model-Driven Engineering, Software Requirements, Domain Models, Gherkin Specifications}

\begin{document}

\begin{abstract}
State-of-the-art Large Language Models (LLMs) excel in code generation at the function level. However, the output quality significantly declines when scaling to repository-level systems. Current workflows relying only on natural language prompts suffer from inherent ambiguity and a lack of verifiability. To address this, we propose \textit{structured spec-driven engineering} (SSDE), a paradigm that leverages structured artifacts to guide LLM generation. We argue that structured specifications as LLM inputs make high-quality, repository-level code generation a tangible goal, while at the same time offering superior verifiability, leading to significant potential for improvement.  We first investigate the feasibility of this vision through a pilot study generating Model-View-Controller (MVC) business logic for three software systems using five LLMs, and then highlight the potential, challenges, and future roadmap for SSDE.
\end{abstract}

\maketitle

\vspace{-0.05in}
\section{Introduction}
\label{sec:introduction}

Recent advances in Large Language Model (LLM)-based code generation and agentic coding assistants motivate new approaches such as \textit{spec-driven engineering}~\cite{github2025specdrivendevelopment}, where evolving specifications guide automated implementation by LLM agents.
These approaches offer significant potential for fundamental changes in software engineering workflows, which traditionally rely on engineers to convert requirements and design to implementation either manually or through bespoke transformations.

\-\hspace{0.25cm} However, while studies show LLMs perform well at generating isolated functions or files~\cite{chen2021humaneval, liu2023evalplus}, their reliability declines for larger or repository-level systems~\cite{saad2025hierarchical, pudari2023copilottopilot}.
In addition, natural language prompts, commonly used in current SDE workflows, are often found to be lossy in communication due to the inherent ambiguity of natural language~\cite{grove2025thenewcode}.
Consequently, this makes it more difficult to generate verification material (e.g., software tests) from the natural language prompts to verify whether LLMs' output adheres to the instructions.
This limitation is fundamental, and simply augmenting the quantity or verbosity of natural language instructions is unlikely to solve the issue, as it does not address the underlying lack of precision required for reliable communication and verification.
We postulate that more structured and maintainable specifications with reduced ambiguity are needed to enable reliable SDE methods.

\-\hspace{0.25cm} For reliable and precise communication, traditional software engineering uses structured specification methods such as Gherkin specifications and software models.
Gherkin is a structured, non-expert-readable language to specify system behavior through verifiable examples~\cite{gherkin2025}, from which engineers can implement software test suites.
Similarly, software models provide automatable and compact representations of software systems, where various types of software models (e.g., domain models, state machines) can capture different software concerns at different levels of abstraction.

\-\hspace{0.25cm} Previous work at the intersection of LLMs and structured specifications remains narrow, with most work focused on using LLMs to generate Gherkin specifications or software models~\cite{chen2023automateddomainmodeling, yang2024iterativedomainmodeling, bozyigit2024domainmodelnlp, silva2025totdomainmodeling}.
Empirical results show that, although using LLMs to automatically produce structured specifications remains largely impractical, LLMs show impressive out-of-the-box ability in inferring software system properties through these specifications~\cite{chen2023automateddomainmodeling}.

\-\hspace{0.25cm} We argue that structured specifications could be the key to consistent and verifiable LLM code generation at system scale, which ultimately aims to step from file-level LLM code generation tasks towards repository-level generation. This paradigm, which we call \textit{structured spec-driven engineering} (SSDE), leverages the structure and verifiability inherent to structured specifications and LLMs' ability to process them for software engineering tasks.

\-\hspace{0.25cm} We hypothesize that structured specifications as LLM input can serve as effective intermediaries between high-level requirements and design, and high-quality repository generation, where specifications (i) effectively guide LLM generation and (ii) offer means to verify the generation outcomes. In this context, LLMs reduce the burden on engineers to manually translate specifications into code, fostering effective human-computer collaboration.

\-\hspace{0.25cm} To investigate the feasibility of our vision, we generate model-view-controller (MVC) ~\cite{syromiatnikov2014mvc} systems in a pilot study, following the SSDE workflow using Gherkin specifications and domain models as inputs.
Our study shows that, with SSDE, high-quality automated repository generation is a tangible goal. The initial results also provide insight into the concrete steps required to realize our vision, offering a roadmap for future research, including steps to improve generation quality and better evaluate the viability of SSDE.

\-\hspace{0.25cm} In the remainder of this vision paper, \Cref{sec:experiment} explains our pilot study, \Cref{sec:results} presents its results, \Cref{sec:roadmap} details our insight gained and our roadmap, followed by the conclusion in~\Cref{sec:conclusion} and the acknowledgement in~\Cref{sec:acknowledgement}.

\vspace{-0.05in}
\section{Experimental Setup}
\label{sec:experiment}

\begin{figure}[tb]
    \centering
    \includegraphics[width=\linewidth]{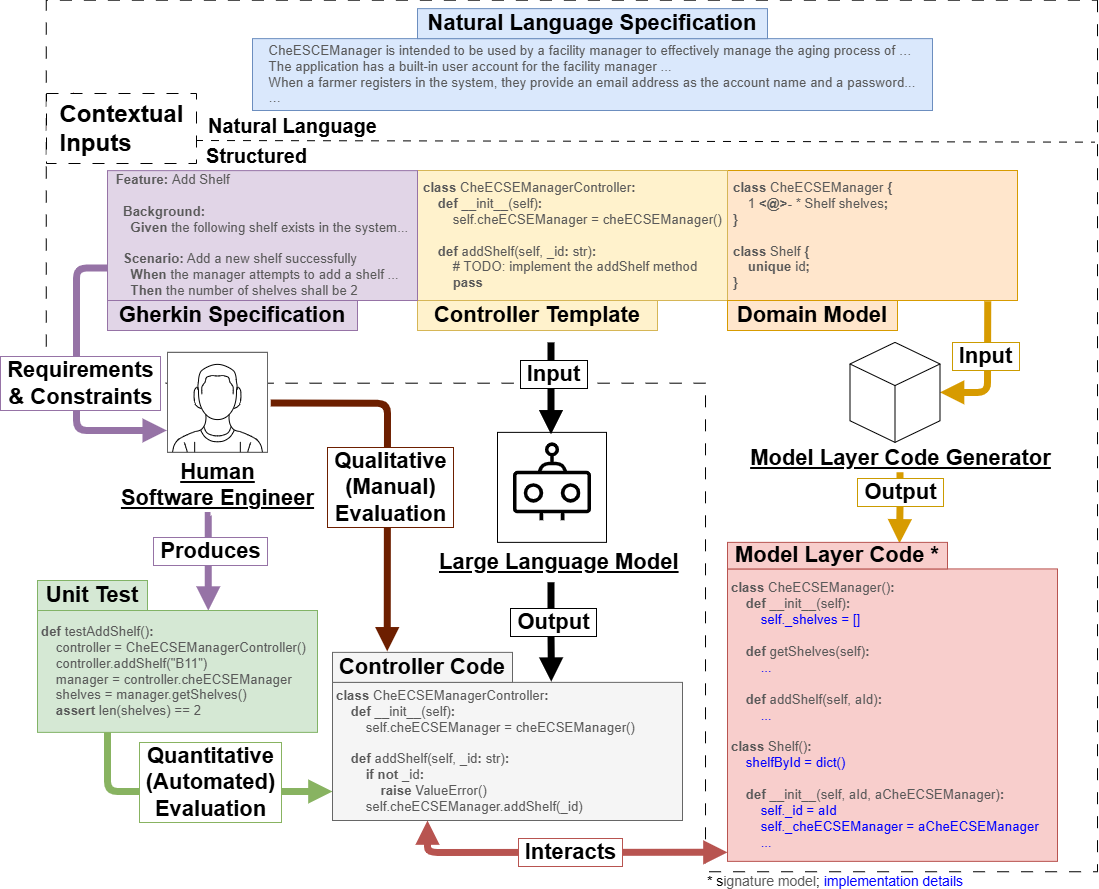}
    \vspace{-0.3in}
    \caption{Overview of SSDE Applied to Pilot Study}
    \label{fig:overview}
    \vspace{-0.1in}
\end{figure}

Our pilot study seeks to demonstrate the feasibility of the \textit{structured spec-driven engineering} (SSDE) workflow and identify challenges for the research roadmap. As detailed in~\Cref{fig:overview}, we use Large Language Models (LLMs) to generate Python MVC business logic (i.e., controllers) for software systems. The generated controller interacts with the model layer to form the full backend system, which is then evaluated for its quality. We use different combinations of inputs for business logic generation and compare the quality of the outcome to understand the contribution of each type of input.

\textbf{Sample Systems.}
We select existing software systems from GitHub. Our inclusion criteria require that each system: (i) supports the Model-View-Controller (MVC) architecture; (ii) contains an up-to-date domain model or equivalent (e.g., PlantUML~\cite{plantuml} diagrams); and (iii) contains an up-to-date Gherkin specification or clearly defined use cases from test suites or design documents. We ultimately selected three systems~\cite{sharifi2020symboleo, cheecsemanager, grzybek2019meetinggroups}, each from a distinct problem domain. As an overview of the systems' complexity, \autoref{tab:system-overview} shows some statistics of these systems' domain models and specifications.

\begin{table}[b]
    \vspace{-0.15in}
    \centering
    \caption{MVC System Overview}
    \label{tab:system-overview}
    \vspace{-0.15in}
    \resizebox{\columnwidth}{!}{
        \begin{tabular}{lcccc}
            \toprule
             & \textbf{Symboleo} & \textbf{CheECSEManager} & \textbf{MeetingGroups} \\
            \midrule
            Num. Classes      & 12  & 17  & 29 \\
            Num. Enum Classes        & 6   & 1   & 12 \\
            Num. Attributes   & 12  & 48  & 88 \\
            Num. Relationships & 26  & 20  & 55 \\
            Num. Test Cases   & 134 & 120 & 119 \\
            \bottomrule
        \end{tabular}
    }
\end{table}

\textbf{Sample LLMs.}
We use five different LLMs for our experiment: Claude Sonnet 4.5 by Anthropic~\cite{anthropic2025claude}, Qwen 3 Coder 480B/A35B Instruct by Qwen~\cite{qwen2025qwen3coder}, GPT 5.1~\cite{openai2025gpt51} and GPT 5 Nano~\cite{openai2025gpt5nano} by OpenAI, and Llama 3.2 3B Instruct by Meta~\cite{meta2024llama32}. The choice of LLMs is primarily based on available state-of-the-art (SOTA) models at the time of the experiment, with a spread of open source (Qwen and Llama), closed source (Claude and GPT), large (Claude, Qwen, and GPT 5.1), and small (GPT 5 Nano and Llama) models.

\-\hspace{0.25cm} \autoref{fig:overview} shows an overview of our Structured Spec-Driven Engineering (SSDE) approach applied to the pilot study.

\textbf{Inputs.}
Besides the controller template, which we always provide as LLM contextual input to guide generation outcome, we compare four types of specification as contextual input as follows.

\-\hspace{0.25cm} \textit{Natural Language Specification ---}
the natural language description of the system's purpose, use cases, and constraints. Though written in free form, this specification is \textit{self-contained}, meaning it includes all essential requirements and constraints for the business logic, such as system behavior and value bounds.

\-\hspace{0.25cm} \textit{Gherkin Specification ---}
standardized scenarios following the systematic template of Gherkin. In contrast to the natural language specifications, the Gherkin specifications contain concrete examples describing expected system behavior under varying conditions, which can be directly mapped to executable test cases.

\-\hspace{0.25cm} \textit{Domain Model ---}
the software model that captures domain concepts and their relationships. Frameworks such as Umple~\cite{lethbridge2021umple} and the Eclipse Modeling Framework (EMF)~\cite{steinberg2008emf} with its Ecore metamodel promote higher levels of abstraction by providing tooling to automatically generate the model layer code within the MVC framework. We consider two types of domain models for the experiment: Umple and Ecore (through \textit{Emfatic}~\cite{garciadominguez2024emfatic}) models. While Umple and Ecore support embedding business logic, we refrain from doing so to evaluate the LLM’s capacity to generate the logic instead.

\-\hspace{0.25cm} \textit{Signature Model ---}
the model layer code’s class and function signatures (API). As shown by the gray text in the red box in~\Cref{fig:overview}, we provide the model layer code's class and function signatures as the model to the LLM. The signatures show the model layer API and calling parameters (which are implicit in domain models as they are generated from the domain model) to help LLM identify functions relevant to the business logic. Umple natively provides support for generating Python code, and Ecore does it through \textit{pyecoregen}~\cite{pyecoregen2021}.

\textbf{Output.}
We prompt the LLM with a combination of selected inputs and ask the LLM to generate the Python business logic that completes the controller template, as shown by the gray box in \Cref{fig:overview}. The generated controller code interacts with the model layer code to form the full backend system.
For our pilot study, we use 0-shot prompting~\cite{kojima2022zeroshot} with no feedback as the naïve approach.

\-\hspace{0.25cm} To reduce LLM's stochastic bias, the business logic generation process is repeated 10 times for each LLM and input configuration evaluated. LLMs use \texttt{0.5} as temperature for Claude, Qwen, and LLama models, or \texttt{medium} as reasoning effort for GPT models, which support the reasoning parameter instead of temperature.

\textbf{Evaluation Method.} We employ both a quantitative benchmark and a qualitative analysis of the generated controller code.

\-\hspace{0.25cm} \textit{Quantitative benchmark.}
Previous studies have shown that generating test cases from Gherkin scenarios is feasible, but not yet flawless~\cite{bergsmann2024automatedexecutiongherkinspecification, poth2025gherkinuitestcase}. For higher reliability, we use a human-made, verified Python unit test suite for our study  as shown by the green box in \Cref{fig:overview}. Each test function corresponds to a specific Gherkin scenario.
During evaluation, we execute the backend system against this test suite. The primary metric for evaluating the generation quality quantitatively is the test pass rate (TPR) from executing the test suite against the generated business logic.

\-\hspace{0.25cm} \textit{Qualitative Analysis.}
We conduct qualitative analysis through manual inspection of the generated business logic across different input configurations, as shown by the brown arrow in \Cref{fig:overview}. The primary goal is to identify recurring failure reasons for LLM-generated business logic and to evaluate code quality attributes missed in the quantitative benchmark.

\vspace{-0.05in}
\section{Results and Analysis}
\label{sec:results}

\begin{table*}[tb]
\centering
\caption{Test Pass Rate of the Generated Business Logic with Claude Sonnet 4.5 using Various Input Configurations}
\label{tab:test-pass-rate}
\vspace{-0.15in}
\begin{tabular}{llccccc}
\toprule
\multirow{2}{*}{\textbf{System}} & \multirow{2}{*}{\textbf{Modeling Tool}} & \multicolumn{3}{c}{\textbf{Natural Language Specification}} & \multicolumn{2}{c}{\textbf{Gherkin Specification}} \\
\cmidrule(lr){3-5} \cmidrule(lr){6-7}
 & & (No Model) & \textbf{Domain Model} & \textbf{Signature Model} & \textbf{Domain Model} & \textbf{Signature Model} \\
\midrule

\multirow{2}{*}{Symboleo} & Umple & 0.0\% ± 0.0\% & 79.9\% ± 0.0\% & 79.9\%± 0.0\%  & \textbf{\underline{99.1\% ± 2.9\%}} & 79.9\% ± 0.0\% \\ \cline{2-7} 
 & Ecore & 47.9\% ± 41.2\% & 79.9\% ± 0.0\% & 79.9\% ± 0.0\% & \textbf{\underline{81.7\% ± 6.1\%}} & 79.9\% ± 0.0\% \\ \hline
\multirow{2}{*}{CheECSEManager} & Umple & 43.5\% ± 3.0\% & 73.0\% ± 3.9\% & 76.7\% ± 0.0\% & 25.7\% ± 7.6\% & \textbf{\underline{79.2\% ± 0.3\%}} \\ \cline{2-7} 
 & Ecore & 16.8\% ± 1.7\% & 53.2\% ± 31.8\% & \textbf{\underline{67.1\% ± 4.0\%}} & 26.0\% ± 18.7\% & 28.2\% ± 19.9\% \\ \hline
\multirow{2}{*}{MeetingGroups} & Umple & 81.6\% ± 0.0\% & 82.2\% ± 6.4\% & \textbf{\underline{85.0\% ± 2.8\%}} & 83.4\% ± 6.2\% & 84.6\% ± 0.7\% \\ \cline{2-7} 
 & Ecore & 31.9\% ± 4.8\% & 82.4\% ± 3.0\% & 78.8\% ± 3.9\% & \textbf{\underline{84.2\% ± 0.9\%}} & 79.8\% ± 0.9\% \\ \hline
\end{tabular}
\end{table*}

\begin{figure}[tb]
    \vspace{-0.15in}
    \centering
    \includegraphics[width=\linewidth]{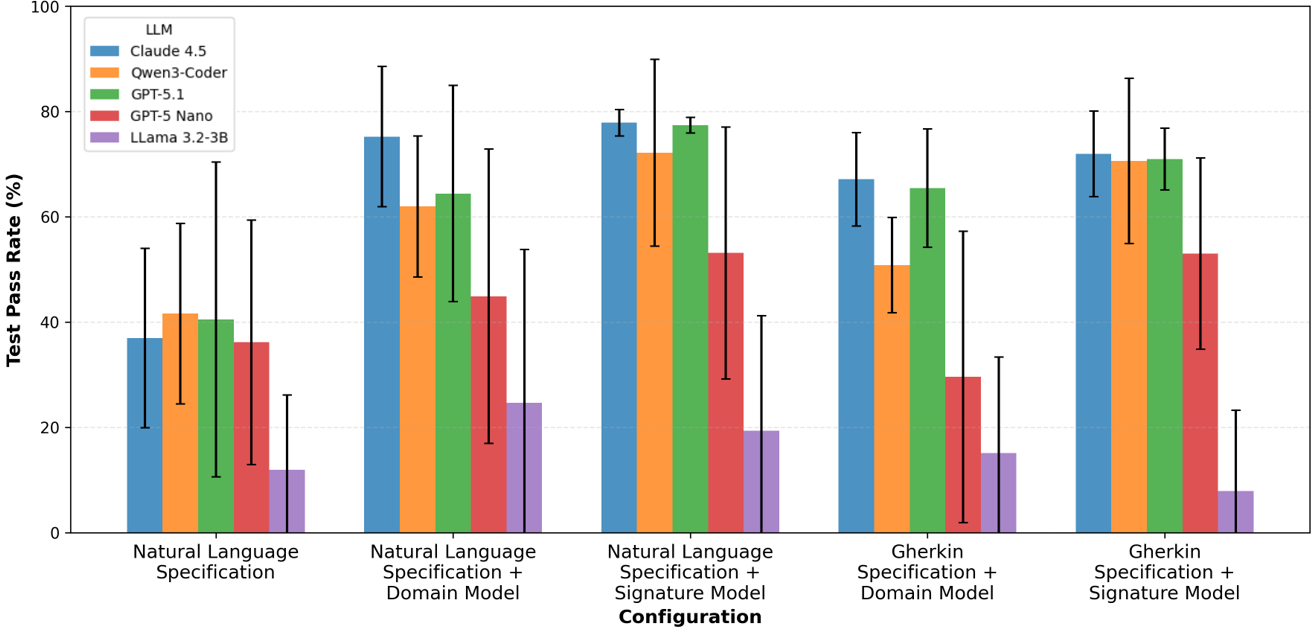}
    \vspace{-0.3in}
    \caption{Overview of the Average Test Pass Rate}
    \label{fig:results_by_config}
\end{figure}

\-\hspace{0.25cm} \autoref{fig:results_by_config} shows the overall test pass rate (TPR) across all systems, modeling tools, and repetitions, grouped by configurations and Large Language Models (LLMs) used.
From the graph, it is clear that adding any type of structured specifications significantly improves output quality compared to the baseline where only the natural language specification is provided.

\-\hspace{0.25cm} Furthermore, while output generated using Gherkin specifications + any model has an overall lower accuracy than the ones generated using natural language specifications + any model (with 6.8\% lower TPR and 0.9\% higher standard deviation averaged across all combinations), the average TPR of Gherkin specifications outputs beats natural language outputs in 14 out of the 30 combinations we tested (5 LLMs × 2 modeling tools × 3 samples), with their average improvement being +7.7\%. This result suggests that there is potential to further improve this approach to surpass natural language in output quality.

\-\hspace{0.25cm} LLMs also show good out-of-the-box ability to infer system specifications through domain models, though providing the model layer’s signatures generated from the domain model instead improves average TPR (+7.82\%) and lowers standard deviation (-2.47\%).

\-\hspace{0.25cm} \autoref{fig:results_by_config} also shows that most LLMs and configurations show high standard deviations in TPR. Manual inspection reveals that these fluctuations are largely attributable to the stochastic nature of LLMs, specifically regarding their ability to avoid common types of errors, which tend to propagate consistently throughout the entire generated controller when any of them manifests, resulting in batches of test failures.
We also experiment with (in addition to the results presented) setting the temperature to 0 for LLMs permitting temperature adjustment, and still find similar trends.

\-\hspace{0.25cm} \autoref{tab:test-pass-rate} shows a closer view of TPR for individual samples, taking Claude as an example, since it is the overall best-performing model, with its highest configuration average TPR over 80\% for 4 out of 6 samples and perfect TPR for several individual runs. From the results, the combination of domain model + Gherkin specifications shows highest average TPR in 3 out of the 6 samples we tested with Claude. Across all LLMs, signature model + natural language specifications shows highest average TPR in 11 out of the 30 LLM-modeling-tool-sample combinations, followed by domain model + Gherkin specifications with 9 out of the 30 combinations.

\-\hspace{0.25cm} However, for both \texttt{CheECSEManager} samples, the domain model and Gherkin specification configuration result in anomalously low average TPR for Claude. Such performance degradations are observed in several configurations across other samples and LLMs, although the specific combinations triggering these outliers vary. Further investigation is required to isolate the underlying causes of these degradations in specific scenarios.

\begin{table}[tb]
\centering
\caption{Distribution of Major Error Types for Failed Tests}
\label{tab:error_distribution}
\vspace{-0.15in}
\begin{tabular}{l|c}
Error Type & \% of All Errors \\ \midrule
Invoking non-existent API & 49.0\% \\
Data type mismatch & 20.2\% \\
Did not validate a given constraint & 11.5\% \\
API positional argument mismatch & 3.2\% \\
Referencing non-existent variables & 1.0\%
\end{tabular}
\vspace{-0.15in}
\end{table}

\-\hspace{0.25cm} \autoref{tab:error_distribution} shows the observed major error types (>1\% of all errors) in the generated controllers that most commonly cause tests to fail. We note that more than 70\% of failures (that is, everything listed except the third row) are caused by errors that can be detected post-generation through static code analysis tools.

\-\hspace{0.25cm} Given these results, we argue that structured specifications have the potential to improve LLM repository-level code generation quality beyond the naïve and the natural language approaches, and that SSDE offers a concrete, viable path towards high-quality, verifiable, repository-level code generation.

\vspace{-0.05in}
\section{Roadmap}
\label{sec:roadmap}

\textbf{Improving Output Quality.}
Our results indicate that while \textit{structured spec-driven engineering} (SSDE) is promising, the approach can be improved in several aspects.

\-\hspace{0.25cm} Incorporating \textit{code analysis} and \textit{feedback loops} is the most immediate path to improving output quality. Our analysis reveals that over 70\% of the failures encountered during the pilot study are detectable through standard static analysis. Since these errors involve structural or type inconsistencies rather than logic flaws, an automated pipeline could parse the abstract syntax tree of the generated code against the model layer and feed these errors back to the LLM~\cite{abtahi2025llmcodeanalysis}. Other types of errors such as constraint validation can also be detected through test scenarios, which can be incorporated into feedback with dynamic analysis~\cite{chen2023teachinglargelanguagemodels}. Given that existing work shows that Large Language Models (LLMs) possess self-correcting capabilities when provided with specific error context~\cite{chen2023teachinglargelanguagemodels, hong2024metagpt, chen2025llm}, this feedback loop could autonomously resolve the bulk of generation issues without human intervention.

\-\hspace{0.25cm} We also observe that although Python is dynamically typed, the \textit{use of type hints} helps provide more guidance to LLMs~\cite{khan2022typerelateddefects, chen2025poweroftypes}. We notice that data type mismatches account for >20\% of test failures during our pilot study. We encourage software automation tools maintainers (e.g., \texttt{umple} and \texttt{pyecoregen}) to include comprehensive type hints in their coding tools, to ensure these frameworks are LLM-ready and optimized for automated engineering workflows.

\-\hspace{0.25cm} There is also potential for \textit{LLM domain knowledge injection} via fine-tuning or advanced prompting to elucidate the mapping between domain model syntax and Python model layer syntax. Although domain models and their corresponding generated model layer code theoretically convey equivalent information, using the latter yields higher test pass rate at the cost of increased input token consumption. This observation suggests that, while current LLMs exhibit proficiency with native Python code, they \textit{lack specific domain knowledge} regarding model code generation tools such as \texttt{umple} and \texttt{pyecoregen}. By equipping LLMs with the ability to infer the expected model layer API directly from compact domain models, we anticipate a further improvement in LLM repository-level generation quality and a substantial reduction in performance fluctuation at lower cost.

\textbf{Evaluating Real-World Impact.}
Assessing the test pass rate reveals only one aspect of SSDE's contribution to the software engineering process. To rigorously evaluate SSDE's practical utility, future work must quantify other benefits of SSDE automation.

\-\hspace{0.25cm} In practice, LLM-enabled engineers often rely on  \textit{tooling support from coding agents} (e.g. GitHub Copilot~\cite{github2021copilot}, Claude Code~\cite{anthropic2025claudecode}) through iterative conversations and feedback to improve code quality, rather than single-shot LLM calls for coding.
To truly assess the viability of SSDE, experiments must extend beyond naïve input-output flow and evaluate the abilities of LLM agentic tools in repository-level code generation.

\-\hspace{0.25cm} We are also actively working towards \textit{a larger, more comprehensive SSDE dataset and benchmark} to evaluate LLM's output quality using different types of inputs with reduced bias in terms of problem domains, variation, and complexity. However, existing software repositories with a complete set of natural language specifications, Gherkin specifications, and domain models are not readily available. While our pilot study uses three such systems obtained from GitHub, advancing the field requires the curation of a larger, more complex dataset to ensure generalizability and to mitigate engineering bias from constructing the inputs for specific experiments. Additionally, future evaluations must expand beyond simple correctness metrics, including the cost-effectiveness of token consumption, sustainability concerns and the output latency associated with different LLMs, behavioural models and input configurations.

\-\hspace{0.25cm} Future research should also quantify the \textit{net productivity gain of SSDE}. Barring groundbreaking innovations, LLM software repository generation will likely remain imperfect in the foreseeable future, necessitating manual intervention by engineers~\cite{becker2025measuring}.
In addition, while structured specifications enable greater automation, their reliance on specific syntax and strict structure imposes an inherent \textit{maintenance cost}. Though we envision some of these tasks to be less burdensome in the future if they can be automated via LLMs~\cite{chen2023automateddomainmodeling, bergsmann2024automatedexecutiongherkinspecification, poth2025gherkinuitestcase}, future research should measure the effort required to produce structured specifications and fix generation errors based on test failures compared to the baseline of manually implementing the system and other automation approaches, such that their difference is the net productivity gain of SSDE in improving efficiency.

\textbf{Applicability.}
SSDE offers a promising approach to extend LLM-automated software engineering to complex software systems and repositories. Still, specific limitations must be addressed to ensure SSDE is easily applicable for general software engineering use.

\-\hspace{0.25cm} First, future research must investigate methods for \textit{partial repository updates} in response to evolving specifications.
Modern software is characterized by continuous evolution, yet due to the high variability of current LLM outputs, there is currently no guarantee that an update on the specification can be accurately reflected on the repository code.
To transform SSDE to a solution for continuous maintenance, it is critical to develop partial code identification and generation techniques~\cite{fruntke2025autofix, omidvar2024humanaicodemigration}, which enables the targeted propagation of specification updates to the existing codebase, ensuring systems can evolve while preserving the value of the existing code.

\-\hspace{0.25cm} Second, given the diverse array of specification languages and software models tailored to specific software concerns, SSDE's role in integrating with a variety of state-of-the-art and future technologies should be explored. As the automation of modeling and specification becomes more accessible, we should investigate how this approach can extend beyond standard software engineering and seek the potential to model complex representations of software systems or even physical objects, enabling LLMs and agentic tools to interpret and interact with the physical world.

\-\hspace{0.25cm} Lastly, other than improving output quality through feedback, the verifiability of structured specifications can also be leveraged to build a variety of practical \textit{software quality tools}. For instance, since Gherkin scenarios are inherently executable, it facilitates the automatic synthesis of \textit{accept tests} and their \textit{harnesses}, a capability we expect to mature as LLMs and software engineering technologies evolve. Furthermore, by treating structured specifications as the ground truth, we can build specialized \textit{static analysis agents} and custom \textit{linters} that validate generated code against the specifications. This ecosystem of tools effectively shifts the engineer's role from code reviewer to specification architect, allowing for a scalable and reliable adoption of repository-level automation.

\vspace{-0.05in}
\section{Conclusion}
\label{sec:conclusion}

We present a vision for moving from file-level Large Langue Model (LLM) coding assistance to repository-level automation through \textit{structured spec-driven engineering} (SSDE).
Rather than relying on purely natural language workflows, which suffer from inherent ambiguity and lack of verifiability, we argue that the path forward lies in leveraging decades of engineering knowledge that led to structured specifications like Gherkin and software models.
Our pilot study generating Model-View-Controller systems shows that structured inputs have the potential to enable high quality repository generation. While challenges remain, we observe great potential to improve the output quality, such as 70\% of errors from the naïve approach can be detected to enable automated corrections. We outline a research roadmap focused on quality, impact, and applicability to improve this approach.
Ultimately, we argue that SSDE offers a promising vision for reducing manual engineering efforts and enabling high-level automation of engineering design.

\vspace{-0.05in}
\section{Acknowledgement}
\label{sec:acknowledgement}

This research was supported by Natural Science and Engineering Research Council of Canada (NSERC) through grant RGPIN-2025-04993.

\bibliographystyle{ACM-Reference-Format}
\bibliography{reference}

@misc{pyecoregen2021,
  author = {Mike Pagel and Vincent Aranega and Andreas Schmidl},
  year = {2021},
  title = {pyecoregen},
  howpublished = {\url{https://github.com/pyecore/pyecoregen/}},
}

@misc{gherkin2025,
  author = {Cucumber},
  title = {Gherkin},
  year = {2025},
  howpublished = {\url{https://github.com/cucumber/gherkin/}},
}

@misc{garciadominguez2024emfatic,
  title = {EMFatic: A textual syntax for EMF Ecore models},
  author = {Antonio García-Domínguez and Dimitris Kolovos},
  howpublished = {\url{https://eclipse.dev/emfatic/}},
  year = {2024},
  organization = {Eclipse Foundation},
}

@article{lethbridge2021umple,
  author = {Timothy C. Lethbridge and others},
  title = {Umple: Model-driven development for open source and education},
  journal = {Science of Computer Programming},
  volume = {208},
  year = {2021},
  issn = {0167-6423},
  doi = {10.1016/j.scico.2021.102665},
}

@book{steinberg2008emf,
  author = {Steinberg, David and Budinsky, Frank and Paternostro, Marcelo and Merks, Ed},
  title = {EMF: Eclipse Modeling Framework 2.0},
  year = {2009},
  isbn = {0321331885},
  publisher = {Addison-Wesley Professional},
  edition = {2nd},
  url = {https://dl.acm.org/doi/10.5555/1197540},
}

@article{chen2021humaneval,
  author = {Mark Chen and others},
  title = {Evaluating Large Language Models Trained on Code},
  journal = {Computing Research Repository},
  year = {2021},
  eprinttype = {arXiv},
  eprint = {2107.03374},
  primaryClass = {cs.LG},
}

@inproceedings{liu2023evalplus,
  title = {Is Your Code Generated by Chat{GPT} Really Correct? Rigorous Evaluation of Large Language Models for Code Generation},
  author = {Jiawei Liu and Chunqiu Steven Xia and Yuyao Wang and Lingming Zhang},
  booktitle = {The 37th Conference on Neural Information Processing Systems (NeurIPS'23)},
  year = {2023},
  url = {https://openreview.net/forum?id=1qvx610Cu7},
}

@inproceedings{chen2023automateddomainmodeling,
  author = {Chen, Kua and Yang, Yujing and Chen, Boqi and Hern\'{a}ndez L\'{o}pez, Jos\'{e} Antonio and Mussbacher, Gunter and Varr\'{o}, D\'{a}niel},
  title = {Automated Domain Modeling with Large Language Models: A Comparative Study},
  year = {2023},
  publisher = {IEEE Press},
  doi = {10.1109/MODELS58315.2023.00037},
  booktitle = {ACM/IEEE 26th International Conference on Model Driven Engineering Languages and Systems (MODELS'23)},
}

@inproceedings{yang2024iterativedomainmodeling,
  author = {Yang, Yujing and Chen, Boqi and Chen, Kua and Mussbacher, Gunter and Varr\'{o}, D\'{a}niel},
  title = {Multi-step Iterative Automated Domain Modeling with Large Language Models},
  year = {2024},
  isbn = {9798400706226},
  publisher = {ACM},
  doi = {10.1145/3652620.3687807},
  booktitle = {Proceedings of the ACM/IEEE 27th International Conference on Model Driven Engineering Languages and Systems (MODELS'24)},
  pages = {587–595},
}

@article{bozyigit2024domainmodelnlp,
  author = {Bozyigit, Fatma and Bardakci, Tolgahan and Khalilipour, Alireza and Challenger, Moharram and Ramackers, Guus and Babur, \"{O}nder and Chaudron, Michel R. V.},
  title = {Generating domain models from natural language text using NLP: a benchmark dataset and experimental comparison of tools},
  year = {2024},
  publisher = {Springer-Verlag},
  volume = {23},
  number = {6},
  issn = {1619-1366},
  doi = {10.1007/s10270-024-01176-y},
  journal = {Software and Systems Modeling},
}

@inproceedings{silva2025totdomainmodeling,
  author = {Silva, Jonathan and Ma, Qin and Cabot, Jordi and Kelsen, Pierre and Proper, Henderik A.},
  title = {Application of the Tree-of-Thoughts Framework to LLM-Enabled Domain Modeling},
  year = {2024},
  isbn = {978-3-031-75871-3},
  publisher = {Springer-Verlag},
  doi = {10.1007/978-3-031-75872-0_6},
  booktitle = {Proceedings of the 43rd International Conference on Conceptual Modeling (ER'24)},
  pages = {94–111},
}

@inproceedings{pudari2023copilottopilot,
  title = {From Copilot to Pilot: Towards AI Supported Software Development},
  author = {Rohith Pudari and Neil A. Ernst},
  year = {2023},
  eprint = {2303.04142},
  archivePrefix = {arXiv},
  primaryClass = {cs.SE},
}

@inproceedings{syromiatnikov2014mvc,
  title = {A journey through the land of model-view-design patterns},
  author = {Syromiatnikov, Artem and Weyns, Danny},
  booktitle = {Proceedings of the 2014 IEEE/IFIP Conference on Software Architecture (ICSA'14)},
  year = {2014},
  organization = {IEEE},
  doi = {10.1109/WICSA.2014.13},
}

@inproceedings{kojima2022zeroshot,
  author = {Kojima, Takeshi and Gu, Shixiang Shane and Reid, Machel and Matsuo, Yutaka and Iwasawa, Yusuke},
  title = {Large language models are zero-shot reasoners},
  year = {2022},
  publisher = {Curran Associates Inc.},
  booktitle = {Proceedings of the 36th International Conference on Neural Information Processing Systems (NeurIPS'22)},
  articleno = {1613, 22199-22213},
  url = {https://dl.acm.org/doi/10.5555/3600270.3601883},
}

@misc{grove2025thenewcode,
  title = {The New Code},
  author = {Sean Grove},
  year = {2025},
  howpublished = {the AI Engineer World's Fair 2025},
  url = {https://www.youtube.com/watch?v=8rABwKRsec4},
}

@misc{github2025specdrivendevelopment,
  title = {Spec-driven development with AI: Get started with a new open source toolkit},
  author = {Den Delimarsky},
  year = {2025},
  howpublished = {GitHub},
  url = {https://github.blog/ai-and-ml/generative-ai/spec-driven-development-with-ai-get-started-with-a-new-open-source-toolkit/},
}

@inproceedings{bergsmann2024automatedexecutiongherkinspecification,
  author = {Bergsmann, Severin and Schmidt, Alexander and Fischer, Stefan and Ramler, Rudolf},
  title = {First Experiments on Automated Execution of Gherkin Test Specifications with Collaborating LLM Agents},
  year = {2024},
  isbn = {9798400711091},
  publisher = {ACM},
  doi = {10.1145/3678719.3685692},
  pages = {12–15},
  booktitle = {Proceedings of the 15th ACM International Workshop on Automating Test Case Design, Selection and Evaluation (A-TEST'24)},
}

@inproceedings{poth2025gherkinuitestcase,
  author = {Poth, Alexander and Rrjolli, Olsi and Wang, Huiyu and Schmid, Klaus},
  editor = {Yilmaz, Murat and Clarke, Paul and Riel, Andreas and Messnarz, Richard and Zelmenis, Mikus and Buce, Ivi Anna},
  title = {Baseline Evaluation of LLM-Facilitated UI Test-Case Generation from Gherkin Specifications},
  booktitle = {Systems, Software and Services Process Improvement},
  year = {2026},
  publisher = {Springer Nature Switzerland},
  isbn = {978-3-032-04288-0},
  doi = {10.1007/978-3-032-04288-0_3},
}

@inproceedings{sharifi2020symboleo,
  author = {Sharifi, Sepehr and Parvizimosaed, Alireza and Amyot, Daniel and Logrippo, Luigi and Mylopoulos, John},
  year = {2020},
  month = {08},
  title = {Symboleo: Towards a Specification Language for Legal Contracts},
  doi = {10.1109/RE48521.2020.00049},
  booktitle = {IEEE 28th International Requirements Engineering Conference (RE'20)},
  note = {Artifact URL: \url{https://github.com/Smart-Contract-Modelling-uOttawa/Symboleo-JS-Core}},
}

@inproceedings{cheecsemanager,
  author = {Padron, Santiago and Valentin, Julien and Olier, Iwan and Séguin, Rémi and Guimond, Mathieu and Shen, Yejia and Yu, Daniel},
  title = {{CheECSEManager}},
  year = {2025},
  publisher = {GitHub},
  howpublished = {\url{https://github.com/F2025-ECSE223/ecse223-group-project-p16}},
}

@misc{grzybek2019meetinggroups,
  author = {Grzybek, Kamil and others},
  title = {{Modular Monolith with DDD}},
  year = {2019},
  publisher = {GitHub},
  howpublished = {\url{https://github.com/kgrzybek/modular-monolith-with-ddd}},
}

@misc{anthropic2025claude,
  title = {Introducing Claude Sonnet 4.5},
  author = {{Anthropic}},
  year = {2025},
  howpublished = {\url{https://www.anthropic.com/news/claude-sonnet-4-5}},
}

@misc{qwen2025qwen3coder,
  title = {Qwen3-Coder},
  author = {{Qwen Team}},
  year = {2025},
  howpublished = {\url{https://github.com/QwenLM/Qwen3-Coder}},
}

@misc{openai2025gpt51,
  title = {GPT-5.1: A smarter, more conversational ChatGPT},
  author = {{OpenAI}},
  year = {2025},
  howpublished = {\url{https://openai.com/index/gpt-5-1/}},
}

@misc{openai2025gpt5nano,
  title = {GPT-5 nano Model},
  author = {{OpenAI}},
  year = {2025},
  howpublished = {\url{https://platform.openai.com/docs/models/gpt-5-nano}},
}

@misc{meta2024llama32,
  title = {llama3.2:3b},
  author = {{Meta AI}},
  year = {2024},
  howpublished = {\url{https://ollama.com/library/llama3.2:3b/}},
}

@inproceedings{chen2025poweroftypes,
  author = {Chen, Boqi and L\'{o}pez, Jos\'{e} Antonio Hern\'{a}ndez and Mussbacher, Gunter and Varr\'{o}, D\'{a}niel},
  title = {The Power of Types: Exploring the Impact of Type Checking on Neural Bug Detection in Dynamically Typed Languages},
  year = {2025},
  publisher = {IEEE Press},
  doi = {10.1109/ICSE55347.2025.00088},
  pages = {489–501},
  booktitle = {Proceedings of the IEEE/ACM 47th International Conference on Software Engineering (ICSE'25)},
}

@article{khan2022typerelateddefects,
  author = {Khan, Faizan and Chen, Boqi and Varro, Daniel and McIntosh, Shane},
  journal = {IEEE Transactions on Software Engineering},
  title = {An Empirical Study of Type-Related Defects in Python Projects},
  year = {2022},
  doi = {10.1109/TSE.2021.3082068},
}

@article{saad2025hierarchical,
  title = {Hierarchical Evaluation of Software Design Capabilities of Large Language Models of Code},
  author = {Saad, Mootez and Chen, Boqi and L{\'o}pez, Jos{\'e} Antonio Hern{\'a}ndez and Varr{\'o}, D{\'a}niel and Sharma, Tushar},
  year = {2025},
  eprint = {2511.20933},
  archivePrefix = {arXiv},
  primaryClass = {cs.SE},
}

@misc{plantuml,
  author = {PlantUML},
  year = {2026},
  title = {PlantUML at a Glance},
  howpublished = {\url{https://plantuml.com/}},
}

@inproceedings{chen2025llm,
  author = {Boqi Chen and Aren A. Babikian and Shuzhao Feng and D{\'{a}}niel Varr{\'{o}} and Gunter Mussbacher},
  title = {LLM-based Satisfiability Checking of String Requirements by Consistent Data and Checker Generation},
  booktitle = {33rd {IEEE} International Requirements Engineering Conference (RE'25)},
  pages = {231--243},
  publisher = {IEEE},
  doi = {10.1109/RE63999.2025.00030},
  year = {2025},
}

@inproceedings{hong2024metagpt,
  title = {Meta{GPT}: Meta Programming for A Multi-Agent Collaborative Framework},
  author = {Sirui Hong and others},
  booktitle = {The 12th International Conference on Learning Representations (ICLR'24)},
  year = {2024},
  url = {https://openreview.net/forum?id=VtmBAGCN7o},
}

@article{becker2025measuring,
  title = {Measuring the Impact of Early-2025 AI on Experienced Open-Source Developer Productivity},
  author = {Joel Becker and Nate Rush and Elizabeth Barnes and David Rein},
  year = {2025},
  eprint = {2507.09089},
  archivePrefix = {arXiv},
  primaryClass = {cs.AI},
}

@article{fruntke2025autofix,
  author = {Fruntke, Lukas and Krinke, Jens},
  title = {Automatically Fixing Dependency Breaking Changes},
  year = {2025},
  doi = {10.1145/3729366},
  journal = {Proceedings of the ACM on Software Engineering},
  articleno = {FSE096},
}

@inproceedings{abtahi2025llmcodeanalysis,
  author = {Abtahi, Seyed Moein and Azim, Akramul},
  booktitle = {IEEE/ACM 2nd International Conference on AI Foundation Models and Software Engineering (FORGE'25)},
  title = {Augmenting Large Language Models with Static Code Analysis for Automated Code Quality Improvements},
  year = {2025},
  doi = {10.1109/Forge66646.2025.00017},
}

@misc{chen2023teachinglargelanguagemodels,
  title = {Teaching Large Language Models to Self-Debug},
  author = {Xinyun Chen and Maxwell Lin and Nathanael Schärli and Denny Zhou},
  year = {2023},
  eprint = {2304.05128},
  archivePrefix = {arXiv},
  primaryClass = {cs.CL},
}

@inproceedings{omidvar2024humanaicodemigration,
  author = {Omidvar Tehrani, Behrooz and M, Ishaani and Anubhai, Anmol},
  title = {Evaluating Human-AI Partnership for LLM-based Code Migration},
  year = {2024},
  isbn = {9798400703317},
  publisher = {ACM},
  doi = {10.1145/3613905.3650896},
  booktitle = {Extended Abstracts of the CHI Conference on Human Factors in Computing Systems (CHI EA'24)},
}

@software{anthropic2025claudecode,
  author = {{Anthropic}},
  title = {Claude Code},
  url = {https://code.claude.com/},
  year = {2025},
}

@software{github2021copilot,
  author = {{GitHub}},
  title = {GitHub Copilot},
  url = {https://github.com/features/copilot},
  year = {2021},
}

\end{document}